# Better Code, Better Sharing:
# On the Need of Analyzing Jupyter Notebooks

Jiawei Wang
Monash University
wangjiawei320@gmail.com

Li Li
Monash University
li.li@monash.edu

Andreas Zeller
CISPA Helmholtz Center for Information Security
zeller@cispa.saarland

*Abstract*—By bringing together code, text, and examples, Jupyter notebooks have become one of the most popular means to produce scientific results in a productive and reproducible way. As many of the notebook authors are experts in their scientific fields, but laymen with respect to software engineering, one may ask questions on the quality of notebooks and their code. In a preliminary study, we experimentally demonstrate that Jupyter notebooks are inundated with poor quality code, e.g., not respecting recommended coding practices, or containing unused variables and deprecated functions. Considering the education nature of Jupyter notebooks, these poor coding practices as well as the lacks of quality control might be propagated into the next generation of developers. Hence, we argue that there is a strong need to programmatically analyze Jupyter notebooks, calling on our community to pay more attention to the reliability of Jupyter notebooks.

## I. INTRODUCTION

Jupyter, a free open-source web application allowing users to write documents containing explanatory text, equations and visualizations, as well as live codes and their execution results, has become tremendously popular nowadays. It provides a way for faculties to make a living workbook for sharing computational information (e.g., code) along with explanations, meanwhile, students can benefit from the live code to better understand the concepts introduced in the notebook. Except for tutoring purposes, Jupyter notebook has also become the data scientists' computational notebook of choice. Indeed, Jupyter has emerged as a de facto standard for data scientists [3]. As argued by Helen Shen on Nature, Jupyter notebook makes data analysis easier to record, understand and reproduce [2].

The popularity of Jupyter notebook is also reflected by the expansion of public Jupyter notebook projects on Github. As of September 2018, there are over *2.5 million Jupyter projects* on Github, which is 10 times more than that of 2015. One main reason contributing to this popularity of Jupyter could be that Jupyter excels in literate programming [8], which allows users to formulate and depict their thoughts with text, supplemented by links, figures, and mathematical equations, as they prepare to write code cells. These code cells can then be executed along with the preparation of the notebook and the results can be permanently recorded, which can further be shared with other users as replicable computational documentation.

Despite the aforementioned benefits, the usage of Jupyter has also come with some drawbacks. As argued by Joel Grus [3], because of inadvertently running code cells out of order, developers may encounter the problem that notebooks do not behave as expected. Moreover, Jupyter notebooks might also encourage poor coding practices, e.g., it is difficult to logically organize the code into a reusable manner. Consider Jupyter notebooks are often used as tutorials or documentation for inexperienced programmers to learn practical programming skills, this poor coding practices may further be propagated into the next generation of developers. This calls for programmatically analyzing Jupyter notebooks—to ensure the quality of the notebooks and the correctness of the code, the consistency between the code and its explanatory text, and more.

By and large, the software engineering community has not yet proposed promising approaches to automatically analyze Jupyter notebooks. To this end, we conduct in this work a preliminary study of a large set of Jupyter notebooks, aiming at checking if the code presented in the notebooks is implemented with good qualities. In a sample of 1982 "high-quality" Jupyter notebooks, our experimental results disclose that the publicly released Jupyter notebooks contain code (1) with poorly respect to the Python style conventions, (2) with unused variables which are defined but never referenced, and (3) accessing deprecated functions.

This preliminary study empirically shows that there is indeed a strong need to analyze Jupyter notebooks. Therefore, based on the experimental results, we further present our vision towards programmatically and systematically analyzing Jupyter notebooks. We argue that our community needs to propose promising approaches to (1) enforce good coding styles, (2) improve the quality and reliability of the code, (3) apply best practices for software quality, and (4) ensure a good balance between text and code in Jupyter notebooks—the more given how many published scientific results depend on calculations made in notebooks.

## II. PRELIMINARY STUDY

Because Jupyter notebooks are frequently used to present tutorials and developer documentation, from which inexperi-enced programmers can learn for certain programming skills, the quality of the notebooks is extremely important. As our initial attempt towards checking the quality of the code written in Jupyter notebooks, we present in this work a preliminary study. Aiming for motivating the need for automated analysis of Jupyter notebooks, we are interested in the following research questions:

**RQ1:** To what extent does the code in Jupyter notebooks respect recommended Python programming conventions?
**RQ2:** Do unused variables exist in Jupyter notebooks?
**RQ3:** Are deprecated library functions used in Jupyter notebooks?

*A. Experimental Setup*

**Dataset.** To answer the aforementioned research questions, we resort to Github to harvest a dataset (i.e., Jupyter notebooks) to support our empirical investigations. Instead of randomly cloning Jupyter notebooks, for which their qualities are unknown, we focus on a set of notable projects that are curated by the Jupyter team. Specifically, the Jupyter team has maintained a gallery of interesting Jupyter notebooks [1]. In this preliminary study, we restrict ourselves on analyzing Python-based notebooks only. After removing dead links and duplicated links, we automatically collected 1,982 notable Python-based notebooks covering various topics such as mathematics, signal processing, natural language processing, etc, as our research subject data.

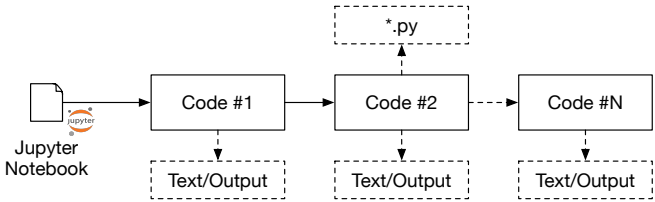

Fig. 1. Preprocessing Jupyter notebooks.

**Preprocessing.** To facilitate the empirical investigation, we develop a set of Python scripts to preprocess the dataset. The programs take a Jupyter notebook file as input and output a chain of code cells. Each code cell is associated with its explanatory text, execution output, and possibly external Python code. The external code is presented as independent Python script (*.py), which is likely written by the same contributors (who write the notebook) and is usually regarded as "library code" to facilitate the implementation of the notebook. Unlike the Python code presented in the notebook, the independent Python scripts will not appear in the notebook but will likely be accessed by the code written in the notebooks.

**Statistics.** The selected 1,982 Jupyter notebooks contain in total 202,332 lines of Python code (LOC). Fig. 2(a) illustrates the distribution of the LOC among the selected projects, giving a median and average lines at 62 and 102.5, respectively. Regarding the number of code cells presented in each notebook, as shown in Fig. 2(b), half of the considered notebooks have presented more than 10 code cells.

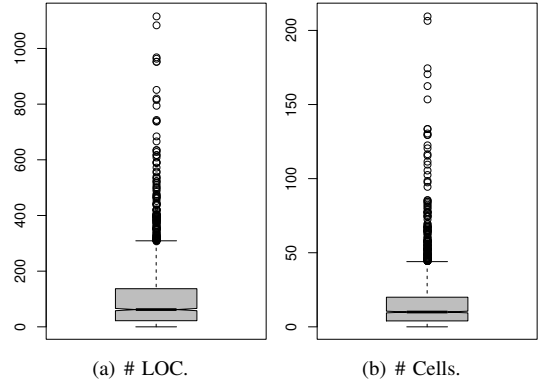

Fig. 2. Distribution of the number of lines of code (left) and the number of code cells (right) in each Jupyter notebook.

*B. RQ1: Python Programming Style*

As the first research question, we are interested in checking if the Python code written in Jupyter notebooks respects the Python coding style. Ideally, because most of the Jupyter notebooks are provided for education, the code should be well aligned with the recommended coding conventions so that the learners will not be misled to write Python code with poor coding practices. To this end, we resort to the *PEP8* checker to evaluate the code written in Jupyter notebooks. This check aims at checking Python code against some of the style conventions in the *PEP8* guidelines, a set of best practices on how to write Python code to improve readability and consistency. The checker takes as input a sequence of Python code and outputs the errors and warnings that the code suffers from.

Among the 1,982 Jupyter notebooks we considered in this work, which correspond to 202,332 lines of Python code in the notebooks, the *PEP8* checker yields 73,371 errors, giving a ratio of 36.26%. This evidence shows that the Python code presented in Jupyter notebooks are not well aligned with the recommended coding practices.

Furthermore, we also launch the checker on all the independent Python scripts (e.g., the external code shown in Fig. 1) located in the same project repository as the Jupyter notebooks are. Among all the related project repositories, 1,919 independent Python scripts are found, corresponding to in total 452,953 lines of code and 60,878 errors given by the *PEP8* checker. The error ratio of independent Python code w.r.t. PEP8 checker is 13.40%, which is much lower than that of code written in the notebooks. This result empirically demonstrates that Jupyter notebook contributors are not attempting to follow good practices while coding. Considering the education nature of Jupyter notebooks, we argue that Jupyter users need to pay more attention to coding practices.

Table I further enumerates the top 10 recurrently appeared error types given by the Python code presented in notebooks and independent Python files. The fact that these two lists are more or less the same suggests that Jupyter notebook contributors are more likely to make mistakes when writing via Jupyter than via independent Python files.

TABLE I
TOP 10 ERROR/WARNING MESSAGES OBSERVED FOR BOTH INDEPENDENT PYTHON SCRIPTS AND NOTEBOOK PYTHON CODE. TO SUPPORT A FAIR COMPARISON, FILE-RELATED MESSAGES (E.G., W292: NO NEWLINE AT END OF FILE) ARE IGNORED.

| Python | remark | Notebook | remark |
|---|---|---|---|
| **E501** | line too long | **E231** | missing whitespace after ,, ;, or : |
| **E231** | missing whitespace after ,, ;, or : | **E501** | line too long |
| **W291** | trailing whitespace | **W293** | blank line contains whitespace |
| **W293** | blank line contains whitespace | **W291** | trailing whitespace |
| E111 | indentation is not a multiple of four | **E225** | missing whitespace around operator |
| E201 | whitespace after ( | **E251** | unexpected spaces around keyword / parameter equals |
| **E265** | block comment should start with # | E703 | statement ends with a semicolon |
| E302 | expected 2 blank lines, found 0 | E261 | at least two spaces before inline comment |
| **E225** | missing whitespace around operator | **E265** | block comment should start with # comment |
| **E251** | unexpected spaces around keyword / parameter equals | E128 | continuation line under-indented for visual indent |

## C. RQ2: Unused Variables

As another experiment towards verifying the quality of the Python code written in Jupyter notebooks, we check in this research question if unused variables are presented by the providers of Jupyter notebooks. Unused variables are such variables that are defined in a code cell but are never used in that cell and its subsequent cells.

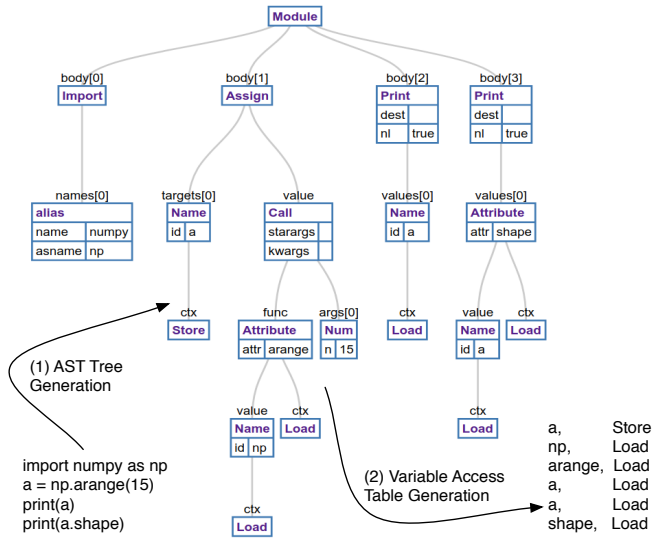

Fig. 3. The working process of identifying unused variables in Python code.

Fig. 3 illustrates the working process of our methodology. Given a piece of Python code, we first build an Abstract syntax tree (AST) for the involving code (as shown in step (1)). Specifically, as shown in the AST, each variable is associated with a special node indicating it is introduced into the context (i.e., Store) or referenced by the context (i.e., Load). In the second step, we perform an in-order traversal over the AST and separate all the variables that are associated with the "Store" or "Load" context into a variable access table. As shown in step (2), the table contains a list of variables following their appearing order in the code. Following this table, if a variable is stored but not loaded subsequently, we will consider it as an unused variable and will flag it as such. If a variable is the result of a cell, e.g. $x = f()$; $x$, the variable (i.e., $x$) will be considered as used (i.e., "Store" in the AST tree).

Following this approach, we experimentally find that 803 notebooks (out of the 1,982) contain code with unused variables. In total, 2,056 unused variables are located. Our manual checking confirms that the reported results are likely correct. We manually verified 20 randomly sampled unused variables, among which all of them are true positive results. This evidence experimentally shows that Jupyter notebooks, even notable ones, are inundated with low-quality code.

## D. RQ3: Deprecated Functions

As the last research question, we are interested in checking if notable Jupyter notebooks contain code accessing into deprecated functions of given libraries. Again, because of the educational purpose, we argue that deprecated functions should be also avoided by notebook contributors.

In order to check whether deprecated functions are used or not, we need to rely on a ground truth of deprecated functions. Unfortunately, such a ground truth is not directly available and usually is non-trivial to infer from the library per se. Fortunately, library maintainer will usually describe the changes (including deprecating some functions) in the release notes of the library. By mining these release notes, one would be able to collect a ground truth of deprecated functions.

Just as a representative example, we manually go through the release notes of the *Scikit-Learn* library published within the past three years (since 2016). *Scikit-Learn* is a popular machine learning library for Python and has been enjoying great popularity in machine learning community. It has been imported by 214 notebooks in our dataset. Table II enumerates the top 5 most used deprecated APIs. Among the 214 notebooks, 75 of them (around 35.05%) have somehow leveraged deprecated APIs (as shown in Table II, one notebook can access multiple deprecated APIs), illustrating that deprecated APIs are quite commonly used by the Jupyter notebook contributors. This evidence once again confirms our previous finding that Jupyter notebooks are inundated with poor quality code. It also suggests that constant maintenance of Jupyter notebooks is also demanded by the community, in order to deliver reliable notebooks to inexperienced learners.

## III. VISION

The aforementioned preliminary study experimentally reveals that Jupyter notebooks, even for notable ones, are inundated with poor coding practices and code smells. Considering the education natural of Jupyter notebooks, the current situation, if not changed, in the long run, would certainly harm the community. The new generations of programmers are educated with poor coding styles that may lead to technical debts, and even with wrong examples that may introduce errors into critical software systems. Therefore, we argue that there is a strong need to properly analyze Jupyter notebooks before releasing them to the public.

TABLE II
TOP FIVE MOST FREQUENTLY DEPRECATED API USES IN SCIKIT-LEARN
INVOLVED NOTEBOOKS

| Deprecated API | # Notebooks | Sample Notebook |
|---|---|---|
| sklearn.cross_validation | 56 | justmarkham/DAT4,herrfz/dataanalysis |
| sklearn.grid_search | 22 | rasbt/pattern_classification |
| sklearn.datasets.fetch_mldata | 6 | jakevdp/PythonDataScienceHandbook |
| sklearn.preprocessing.Imputer | 3 | ogrisel/parallel_ml_tutorial |
| sklearn.mixture.GMM | 3 | jakevdp/PythonDataScienceHandbook |

We now enumerate some of the future directions that are needed to be addressed by the community.

**Enforcing good coding styles.** The fact that notable Jupyter notebooks have their code written without fair respect to the recommended coding conventions suggests that there is no attempt yet for regulating Jupyter users to write source code with good programming styles. However, poor coding styles can be learned and hence propagated into thousands of programmers who might write more code with poor coding practices. Therefore, we argue that our community should implement effective approaches to enforce good coding styles in Jupyter notebooks.

**Improving code quality and reliability.** Automated tools are expected to locate poor quality code (or code smells) and subsequently to recommend fixes to improve the code quality, so as to improve the overall quality of Jupyter notebooks. In addition to the occurrences of unused variables, the access of deprecated functions, many other topics (such as the usage of duplicated code and inefficient algorithms, etc.) are also worth to explore.

**Apply best practices for software quality.** The Software Engineering community has produced a wealth of best practices to ensure software quality. Like other software, Jupyter Notebooks can be tested, verified, reviewed, assessed. Users of Jupyter Notebooks should be encouraged to apply static checkers and bug finders; use tests and assertions for systematic checks; provide specifications on result properties; and use and apply domain-specific consistency checks for Notebook results. This also calls for better tools that analyze and check Notebook code—including static analysis for Python, Julia, or R code.

**Ensure a good balance between text and code.** Jupyter notebooks embrace an innovative way of sharing knowledge, where the intricacies are not only explained but also complemented with live coding examples. However, too much water can drown the miller. We argue that a good balance between the explanatory text and the code is preferred. The flow of the code and text should be also kept consistent. To achieve this purpose, we believe that an interdisciplinary approach, which involves code analysis and comprehension, natural language processing, and artificial intelligence, could be highly useful.

## IV. RELATED WORK

To the best of our knowledge, our work is the first investigation motivating the necessity of deep static/dynamic analysis of Jupyter notebooks—a requirement widely overlooked by the software research community. Indeed, the only work targeting the analysis of Jupyter notebooks, for which that we are aware of, is the one recently conducted by Pimentel et al. [10], who empirically look at the reproducibility of Jupyter notebooks. Their experimental results show that the success rate of reproducing Python notebooks are quite low (less than 25%). This evidence further supplements our initiative towards calling on our community to propose advanced approaches for analyzing Jupyter notebooks.

Jupyter notebooks have been popularly investigated by our fellow researchers of other communities [7], [5], [4], [11], [6], [9]. For example, Rule et al. [11] look at computational notebooks from the human factors point of view. Based on a large scale empirical study of computational notebooks on Github, the authors show that not all computational notebooks contain explanatory text and only a small set of notebooks have discussed the reasoning or results of the methods described. Through an interview with 15 academic data analysts, they argue that computational notebooks are considered to be messy. These results complement our work and demonstrate the necessity of analyzing Jupyter notebooks.

## V. CONCLUSION

In this work, we conducted a preliminary study on a set of notable Jupyter notebooks. Our experimental results reveal that Jupyter notebooks are indeed inundated with poor coding practices. Motivated by these empirical results, we presented our vision on the need of analyzing Jupyter notebooks, appealing for the software engineering community to pay more attention to the quality and reliability of Jupyter notebooks.

REFERENCES

[1] A gallery of interesting Jupyter Notebooks. https://github.com/jupyter/jupyter/wiki/A-gallery-of-interesting-Jupyter-Notebooks. Accessed: 2019-05-10.
[2] Interactive notebooks: Sharing the code. https://www.nature.com/news/interactive-notebooks-sharing-the-code-1.16261. Accessed: 2019-05-10.
[3] Why Jupyter is data scientists' computational notebook of choice. https://www.nature.com/articles/d41586-018-07196-1. Accessed: 2019-05-10.
[4] Björn A Grüning, Eric Rasche, Boris Rebolledo-Jaramillo, Carl Eberhard, Torsten Houwaart, John Chilton, Nate Coraor, Rolf Backofen, James Taylor, and Anton Nekrutenko. Jupyter and galaxy: easing entry barriers into complex data analyses for biomedical researchers. *PLoS computational biology*, 13(5):e1005425, 2017.
[5] Jessica B Hamrick. Creating and grading ipython/jupyter notebook assignments with nbgrader. In *Proceedings of the 47th ACM Technical Symposium on Computing Science Education*, pages 242–242. ACM, 2016.
[6] Mary Beth Kery and Brad A Myers. Interactions for untangling messy history in a computational notebook. In *2018 IEEE Symposium on Visual Languages and Human-Centric Computing (VL/HCC)*, pages 147–155. IEEE, 2018.

[7] Thomas Kluyver, Benjamin Ragan-Kelley, Fernando Pérez, Brian E Granger, Matthias Bussonnier, Jonathan Frederic, Kyle Kelley, Jessica B Hamrick, Jason Grout, Sylvain Corlay, et al. Jupyter notebooks-a publishing format for reproducible computational workflows. In *ELPUB*, pages 87–90, 2016.

[8] Donald Ervin Knuth. Literate programming. *The Computer Journal*, 27(2):97–111, 1984.

[9] Neglectos. A Preliminary Analysis on the Use of Python Notebooks. 2018.

[10] Joao Pimentel, Leonardo Murta, Vanessa Braganholo, and Juliana Freire. A large-scale study about quality and reproducibility of jupyter notebooks. 2019.

[11] Adam Rule, Aurélien Tabard, and James D Hollan. Exploration and explanation in computational notebooks. In *Proceedings of the 2018 CHI Conference on Human Factors in Computing Systems*, page 32. ACM, 2018.